\documentclass[12pt,english]{article}
\usepackage[T1]{fontenc}
\usepackage[latin1]{inputenc}
\usepackage{geometry}
\usepackage{graphicx}

\makeatletter

\usepackage{babel}
\makeatother
\begin{document}

\smallskip{}
\title{Quantum Critical Point and Entanglement in a Matrix Product Ground
State}

\author{Amit Tribedi and Indrani Bose }

\maketitle
\begin{center}Department of Physics\end{center}

\begin{center}Bose Institute\end{center}

\begin{center}93/1, Acharya Prafulla Chandra Road\end{center}

\begin{center}Kolkata - 700 009, India\end{center}

\begin{abstract}
In this paper, we study the entanglement properties of a spin-1 model
the exact ground state of which is given by a Matrix Product state.
The model exhibits a critical point transition at a parameter value
$a=0$. The longitudinal and transverse correlation lengths are known
to diverge as $a\rightarrow0$. We use three different entanglement
measures $S(i)$ (the one-site von Neumann entropy), $S(i,j)$ (the
two-body entanglement) and $G(2,n)$ (the generalized global entanglement)
to determine the entanglement content of the MP ground state as the
parameter $a$ is varied. The entanglement length, associated with
$S(i,j)$, is found to diverge in the vicinity of the quantum critical
point $a=0$. The first derivative of the entanglement measure $E$
$(=S(i),\: S(i,j))$ w.r.t. the parameter $a$ also diverges. The
first derivative of $G(2,n)$ w.r.t. $a$ does not diverge as $a\rightarrow0$
but attains a maximum value at $a=0$. At the QCP itself all the three
entanglement measures become zero. We further show that multipartite
correlations are involved in the QPT at $a=0$.
\end{abstract}

\section*{I. INTRODUCTION}

\smallskip{}
The entanglement characteristics of the ground states of many body
Hamiltonians describing condensed matter systems constitute an important
area of study in quantum information theory. Entanglement is an essential
resource in quatum computation and communication protocols. Condensed
matter, specially, spin systems have been proposed as candidate systems
for the realization of some of the protocols. Entanglement provides
a measure of non-local quantum correlations in the system and it is
of significant interest to determine how the correlations associated
with the ground state of the system change as one or more than one
parameter of the system is changed. The focus on ground state characteristics
arises from the possibility of quantum phase transitions (QPTs) which
occur at temperature $T=0$ (when the system is in its ground state)
and are driven solely by quantum fluctuations \cite{key-1}. A QPT
is brought about by tuning a parameter, either external or intrinsic
to the Hamiltonian, to a special value termed the transition point.
In thermodynamic critical phenomena, the thermal correlation length
diverges and the thermodynamic quantities become singular as the critical
point is approached. In the quantum case, the correlation length diverges
in the vicinity of the QCP and the ground state properties develop
non-analytic features. An issue of considerable interest is whether
the quantum correlations, like the usual correlation functions, become
long-ranged near the QCP. In a wider perspective, the major goal is
to acquire a clear understanding of the variation in entanglement
characteristics as a tuning parameter is changed. QPTs have been extensively
studied in spin systems both theoretically and experimentally. In
recent years, several theoretical studies have been undertaken to
elucidate the relationship between QPTs and entanglement in spin systems
\cite{key-2,key-3,key-4,key-5,key-6,key-7}. In particular, a number
of entanglement measures have been identified which develop special
features close to the transition point. One such measure is concurrence
which quantifies the entanglement between two spins $(S=\frac{1}{2})$.
At a QCP, as illustrated by a class of exactly-solvable spin models
$(S=\frac{1}{2})$, the derivative of the ground state concurrence
has a logarithmic singularity though the concurrence itself is non-vanishing
upto only next-nearest-neighbour-distances between two spins \cite{key-2,key-3}.
Discontinuities in the ground state concurrence have been shown to
characterize first order QPTs \cite{key-8,key-9,key-10}. Later, Wu
et al. \cite{key-5} showed that under some general assumptions a
first order QPT, associated with a discontinuity in the first derivative
of the ground state energy, gives rise to a discontinuity in a bipartite
entanglement measure like concurrence and negativity. Similarly, a
discontinuity or a divergence in the first derivative of the same
entanglement measure is the signature of a second order phase transition
with a discontinuity or a divergence in the second derivative of the
ground state energy. Another measure of entanglement, studied in the
context of QPTs, is the entropy of entanglement between a block of
$L$ adjacent spins in a chain with the rest of the system \cite{key-4}.
At the QCP, the entropy of entanglement diverges logarithmically with
the length of the block. There is, however, no direct relation with
the long range correlations in the system.

A number of entanglement measures have recently been proposed which
are characterized by a diverging length scale, the entanglement length,
close to a QCP. The localizable entanglement (LE) between two spins
is defined as the maximum average entanglement that can be localized
between them by performing local measurements on the rest of the spins
\cite{key-11}. The entanglement length sets the scale over which
the LE decays. The two-body entanglement $S(i,j)$ is a measure of
the entanglement between two separated spins, at sites $i$ and $j$,
and the rest of the spins \cite{key-7}. Let $\rho(i,j)$ be the reduced
density matrix for the two spins, obtained from the full density matrix
by tracing out the spins other than the ones at sites $i$ and $j$.
The two body entanglement $S(i,j)$ is given by the von Neumann entropy

\begin{equation}
S(i,j)=-Tr\,\rho(i,j)\, log_{2}\,\rho(i,j)\label{1}\end{equation}
In a translationally invariant system, $S$ depends only on the distance
$n=\mid j-i\mid$. As pointed out in \cite{key-7}, the spins that
are entangled with one or both the spins at sites $i$ and $j$ contribute
to $S$. The following results have been obtained in the case of the
$S=\frac{1}{2}$ exactly solvable anisotropic $XY$ model in a transverse
magnetic field. The model, away from the isotropic limit, belongs
to the universality class of the transverse Ising model. The two-body
entanglement $S(i,j)$ has a simple dependence on the spin correlation
functions in the large $n$ limit. Away from the critical point, $S(i,j)$
is found to saturate over a length scale $\xi_{E}$ as $n$ increases.
Near the QCP, one obtains

\begin{equation}
S(i,j)-S(\infty)\sim n^{-1}\, e^{-\frac{n}{\xi_{E}}}\label{2}\end{equation}
The entanglement length (EL), $\xi_{E}$, has an interpretation similar
to that in the case of LE. The EL diverges with the same critical
exponent as the correlation length at the QCP. $S(i,j)$ thus captures
the long range correlations associated with a QPT. At the critical
point itself, $S(i,j)-S(\infty)$ has a power-law decay, i.e., $S(i,j)-S(\infty)\sim n^{-\frac{1}{2}}$.
In the limit of large $n$, the first derivative of $S(i,j)$ w.r.t.
a Hamiltonian parameter develops a $\lambda-$like cusp at the critical
point. The universality and a finite-size scaling of the entanglement
have also been demonstrated. The one-site von Neumann entropy

\begin{equation}
S(i)=-Tr\,\rho(i)\, log_{2}\,\rho(i)\label{3}\end{equation}
is also known to be a good indicator of a QPT \cite{key-3}. It provides
a measure of how a single spin at the site $i$ is entangled with
the rest of the system. The reduced density matrix $\rho(i)$ is obtained
from the full density matrix by tracing out all the spins except the
one at the site $i$. Oliveira et al. \cite{key-6} have proposed
a generalized global entanglement (GGE) measure $G(2,n)$ which quantifies
multipartite entanglement (ME). $G(2,n)$ for a translationally symmetric
system is given by

\begin{equation}
G(2,n)=\frac{d}{d-1}[1-\sum_{l,m=1}^{d^{2}}\mid[\rho(j,j+n)]_{lm}\mid^{2}]\label{4}\end{equation}
where $\rho(j,j+n)$ is the reduced density matrix of dimension $d$.
The factor 2 in $G(2,n)$ indicates that the reduced density matrix
is that for a pair of particles. Wu et al. \cite{key-5} considered
QPTs characterized by non-analyticities in the derivatives of the
ground state energy. These arise from the non-analyticities in one
or more of the elements of the reduced density matrix. In terms of
the GGE, a discontinuity in $G(2,n)$ signals a first order QPT, brought
about by a discontinuity in one or more of the elements, $[\rho_{j,j+n}]_{lm}$
of the reduced density matrix \cite{key-6}. A discontinuity or divergence
in the first derivative of $G(2,n)$ w.r.t. the tuning parameter occurs
due to a discontinuity or divergence in the first derivetives of one
or more of the elements of the reduced density matrix. The associated
QPT is of second order. Non-analyticities in $G(2,n)$ and its derivatives
thus serve as indicators of QPTs. In the case of the $XY$ $S=\frac{1}{2}$
spin chain, the GGE measure shows a diverging EL as the QCP is approached.
The EL $\xi_{E}=\frac{\xi_{C}}{2}$ where $\xi_{C}$ is the usual
correlation length. Thus, both the length scales diverge with the
same critical exponent near the QCP.

The relationship between entanglement and QPTs has mostly been explored
for spin-$\frac{1}{2}$ systems. The entanglement properties of the
ground states of certain spin$-1$ Hamiltonians have been studied
using different measures \cite{key-11,key-12,key-13}. Numerical studies
show that the LE has the maximal value for the ground state of the
spin-1 Heisenberg antiferromagnet with open boundary conditions (OBC)
\cite{key-13}. In the case of the spin-1 Affleck-Kennedy-Lieb-Tasaki
(AKLT) model \cite{key-14}, the result can be proved exactly. A class
of spin-1 models, the $\phi$-deformed AKLT models, is characterized
by an exponentially decaying LE with a finite EL $\xi_{E}$. The length
$\xi_{E}$ diverges at the point $\phi=0$ though the conventional
correlation length remains finite \cite{key-13}. A recent study \cite{key-15}
shows that in the case of spin-1 systems, the use of LE for the detection
of QPTs is not feasible. An example is given by the $S=1$ XXZ Heisenberg
antiferromagnet with single-ion anisotropy. The model has a rich phase
diagram with six different phases. The LE is found to be always 1
in the entire parameter region and hence is insensitive to QPTs. The
ground states of certain spin-1 models have an exact representation
in terms of matrix product states (MPS) \cite{key-16,key-17,key-18}.
The ground state of the spin-1 AKLT model, termed a valence bond solid
(VBS) state, is an example of an MPS. The ground state is characterized
by short-ranged spin-spin correlations and a hidden topological order
known as the string order. The excitation spectrum of the model is
further gapped. In the MPS formalism, ground state expectation values
like the correlation functions are easy to calculate. This has made
it particularly convenient to study phase transitions in spin models
with MP states as exact ground states \cite{key-17}. The transitions
identified so far include both first and second order transitions
and are brought about by the tuning of the Hamiltonian parameters.
The second order transition in the class of finitely correlated MP
states, however, differs from the conventional QPT in one important
respect. The spin correlation function is always of the form $A_{C}\, e^{-\frac{n}{\xi_{C}}}$
for large $n$. The correlation length $\xi_{C}$ diverges as the
transition point is approached. The pre-factor $A_{C}$, however,
vanishes at the transition point \cite{key-17}. This is in contrast
to the power-law decay of the correlation function at a conventional
QCP. Some distinct features of QPTs in MP states have recently been
identified \cite{key-19}. One of these relates to the analyticity
of the ground state energy density for all values of the tuning parameter.
In a conventional QPT, the energy density becomes non-analytic at
the QCP. The MP states appear to provide an ideal playground for exploring
novel types of QPTs. In this paper, we consider a spin-1 model, the
exact ground state of which is given by an MP state \cite{key-20}.
The model has a rich phase diagram with a number of first order phase
transitions and a critical point transition. We study the entanglement
properties of the ground state with a view to pinpoint the special
features which appear close to the critical point. This is done by
using three different entanglement measures, namely, the single-site,
two-body and generalized global entanglement defined earlier.

\section*{II. REDUCED DENSITY MATRIX OF MP }

\section*{GROUND STATE}

We consider a spin-1 chain Hamiltonian proposed by Kl\"{u}mper et
al. \cite{key-20} which describes a large class of antiferromagnetic
(AFM) spin-1 chains with MP states as exact ground states. The Hamiltonian
satisfies the symmetries : (i) rotational invariance in the $x-y$
plane, (ii) invariance under $S^{z}\rightarrow-S^{z}$ and (iii) translation
and parity invariance. The Hamiltonian has the general form

\[
H=\sum_{j=1}^{L}h_{j,\, j+1}\]

\[
h_{j,\, j+1}=\alpha_{0}A_{j}^{2}+\alpha_{1}(A_{j}B_{j}+B_{j}A_{j})+\alpha_{2}B_{j}^{2}+\alpha_{3}A_{j}+\alpha_{4}B_{j}(1+B_{j})+\]

\begin{equation}
+\alpha_{5}((S_{j}^{z})^{2}+(S_{j+1}^{z})^{2}+C\label{5}\end{equation}
where $L$ is the number of sites in the chain and periodic boundary
conditions (PBC) hold true. The parameters $\alpha_{j}$ are real
and $C$ is a constant. The nearest-neighbour (n.n.) interactions
are

\begin{equation}
\begin{array}{c}
A_{j}=S_{j}^{x}S_{j+1}^{x}+S_{j}^{y}S_{j+1}^{y}\\
B_{j}=S_{j}^{z}S_{j+1}^{z}\end{array}\label{6}\end{equation}
The constant $C$ in Eq. (5) may be adjusted so that the ground state
eigenvalue of $h_{j,\, j+1}=0$. Hence

\begin{equation}
h_{j,\, j+1}\geq0\quad\Rightarrow H\geq0\label{7}\end{equation}
 i.e., $H$ has only non-negative eigenvalues. In the AFM case, the
$z$-component of the total spin of the ground state $S_{tot}^{z}=0$.
Kl\"{u}mper et al. showed that in a certain subspace of the $\alpha_{j}-$parameter
space , the AFM ground state has the MP form. Let $\left|0\right\rangle $
and $\left|\pm\right\rangle $ be the eigenstates of $S^{z}$ with
eigenvalues $0$, $+1$ and $-1$ respectively. Define a $2\times2$
matrix at each site $j$ by

\begin{equation}
g_{j}=\left(\begin{array}{cc}
\left|0\right\rangle  & -\sqrt{a}\left|+\right\rangle \\
\sqrt{a}\left|-\right\rangle  & -\sigma\left|0\right\rangle \end{array}\right)\label{8}\end{equation}
with non-vanishing parameters $a,\,\sigma\neq0$.

The global AFM state is written as

\begin{equation}
\left|\psi_{0}\,(a,\,\sigma)\right\rangle =Tr\,(g_{1}\otimes g_{2}\otimes......\otimes g_{L})\label{9}\end{equation}
where `$\otimes$' denotes a tensor product. One can easily check
that $S_{tot}^{z}\left|\psi_{0}\right\rangle =0$, i.e., the state
is AFM. One now demands that the state $\left|\psi_{0}\,(a,\,\sigma)\right\rangle $
is the exact ground state of the Hamiltonian $H$ with eigenvalues
$0$. For this, it is sufficient to show that

\begin{equation}
h_{j,\, j+1}\:(g_{j}\otimes g_{j+1})=0\label{10}\end{equation}

Eq. (3) and (10) are satisfied provided the following equalities

\begin{equation}
\begin{array}{cc}
1)\,\sigma=sign(\alpha_{3}), & 2)\, a\,\alpha_{0}=\alpha_{3}-\alpha_{1},\\
3)\,\alpha_{5}=\mid\alpha_{3}\mid+\alpha_{0}(1-a^{2}), & 4)\,\alpha_{2}=\alpha_{0}a^{2}-2\mid\alpha\mid\end{array}\label{11}\end{equation}
and inequalities

\begin{equation}
a\neq0,\:\alpha_{3}\neq0,\:\alpha_{4}>0,\:\alpha_{0}>0\label{12}\end{equation}
hold true. The state $\left|\psi_{0}\,(a,\,\sigma)\right\rangle $
is the ground state of the Hamiltonian $(5)$ with ground state energy
zero provided the equalities in $(8)$ are satisfied. The inequalities
constrain the other eigenvalues of $h_{j,j+1}$ to be positive. If
the inequalities are satisfied, the ground state can be shown to be
unique for any chain length $L$. Also, in the thermodynamic limit
$L\rightarrow\infty$, the excitation spectrum has a gap $\Delta$.
With equality signs in the inequalities $(12)$, the state $\left|\psi_{0}\,(a,\,\sigma)\right\rangle $
is still the ground state but is no longer unique. The spin$-1$ model
has the typical feature of a Haldane-gap (HG) antiferromagnet. In
fact, the AKLT model is recovered as a special case with $a=2,\:\sigma=1,\:\alpha_{3}=3\alpha_{0}>0,\:\alpha_{2}=-2\alpha_{0}$
and $\alpha_{4}=3\alpha_{0}$. The state $(9)$ now represents the
VBS state.

Using the transfer matrix method \cite{key-16}, the ground state
correlation functions can be calculated in a straightforward manner.
The results are $(L\rightarrow\infty,\, r\geq2)$ :

\noindent Longitudinal correlation function 

\begin{equation}
\left\langle S_{1}^{z}\, S_{r}^{z}\right\rangle =-\frac{a^{2}}{(1-|a|)^{2}}\left(\frac{1-|a|}{1+|a|}\right)^{r}\label{13}\end{equation}
 Transverse correlation function\begin{equation}
\left\langle S_{1}^{x}\, S_{r}^{x}\right\rangle =-|a|\,[\sigma+sign\, a]\left(\frac{-\sigma}{1+|a|}\right)^{r}\label{14}\end{equation}
The correlations $(13)$ and $(14)$ decay exponentially with the
longitudinal and transverse correlation lengths given by\begin{equation}
\xi_{l}^{-1}=ln\left|\frac{1+|a|}{1-|a|}\right|,\;\;\xi_{t}^{-1}=ln(1+|a|)\label{15}\end{equation}
Furthermore, the string order parameter has a non-zero expectation
value in the ground state. One finds that the correlation lengths
diverge as $a\rightarrow0$. At the point $a=0$, the correlation
functions given by Eq. $(13)$ and $(14)$ are zero. At a conventional
QCP, the correlation functions have a power-law decay. We will, however,
refer to the point as a QCP since the correlation lengths diverge
as the point is approached. A consequence of the diverging correlation
length is that the excitation spectrum of the spin-1 model, which
is gapped (the Haldane phase) for $a>0$, becomes gapless at the critical
point $a=0$ \cite{key-20}. The presence or absence of a gap in the
excitation spectrum of a system is reflected in the low temperature
thermodynamic properties of the system. Furthermore, the string order
parameter has a non-zero expectation value in the ground state for
$a>0$ and becomes zero at $a=0$ indicating the appearance of a new
phase. Refs. \cite{key-16,key-17} provide several other examples
of spin-1 models with finitely correlated MP states as exact ground
states. All these models exhibit critical point transitions with features
similar to those in the case of the spin-1 model described by the
Hamiltonian in Eq. $(5)$. We now focus on the entanglement properties
of the MP ground state (Eq. $(9)$). We consider $a$ to be $\geq0$
and $\sigma=+1$ in Eq. $(8)$. The one-site reduced density matrix
$\rho(i)$ (Eq. $(3)$) obtained by tracing out all the spins except
the $i$-th spin from the ground state density matrix $\rho=\left|\psi_{0}\right\rangle $$\left\langle \psi_{0}\right|$,
can be calculated using the transfer matrix method \cite{key-16}.
The density matrix, from Eq. $(9)$, is\begin{equation}
\rho=\left|\psi_{0}\right\rangle \left\langle \psi_{0}\right|=\sum_{\{ n_{\alpha},m_{\alpha}\}}g_{n_{1}n_{2}}\, g_{n_{2}n_{3}}........g_{n_{L}n_{1}}\: g_{m_{1}m_{2}}^{\dagger}\, g_{m_{2}m_{3}}^{\dagger}......g_{m_{L}m_{1}}^{\dagger}\label{16}\end{equation}
The summation is over all the indices, $n_{i}$, $m_{i}$, $i=1,2,.....L$. 

We define a $4\times4$ matrix $f$ (the elements of which are operators)
at any lattice site as \begin{equation}
f_{\mu_{1}\mu_{2}}\Rightarrow f_{(n_{1},m_{1})(n_{2},m_{2})}\equiv g_{n_{1}n_{2}}\, g_{m_{1}m_{2}}^{\dagger}\label{17}\end{equation}
The convention of the ordering of the multi-indices is $\mu=1,2,3,4\:\leftrightarrow(11),(12),(21),(22)$.
Thus, $f$ can be written as \begin{equation}
f=\left(\begin{array}{cccc}
\left|0\right\rangle \left\langle 0\right| & -\sqrt{a}\left|0\right\rangle \left\langle 1\right| & -\sqrt{a}\left|1\right\rangle \left\langle 0\right| & a\left|1\right\rangle \left\langle 1\right|\\
\sqrt{2}\left|0\right\rangle \left\langle -1\right| & -\left|0\right\rangle \left\langle 0\right| & -a\left|1\right\rangle \left\langle -1\right| & \sqrt{a}\left|1\right\rangle \left\langle 0\right|\\
\sqrt{a}\left|-1\right\rangle \left\langle 0\right| & -a\left|-1\right\rangle \left\langle 1\right| & -\left|0\right\rangle \left\langle 0\right| & \sqrt{a}\left|0\right\rangle \left\langle 1\right|\\
a\left|-1\right\rangle \left\langle -1\right| & -\sqrt{a}\left|-1\right\rangle \left\langle 0\right| & -\sqrt{a}\left|0\right\rangle \left\langle -1\right| & \left|0\right\rangle \left\langle 0\right|\end{array}\right)\label{18}\end{equation}
Also,\begin{equation}
\rho(i)=Tr_{1,..L}^{i}\left|\psi_{0}\right\rangle \left\langle \psi_{0}\right|\label{19}\end{equation}
where the trace is over all the spins except the $i$-th one. The
transfer matrix $F$ at a site $m$ is obtained by taking the trace
over $f$ at the same site, i.e., \begin{equation}
F_{m}=\sum_{k}\left\langle k\right|f_{m}\left|k\right\rangle \label{20}\end{equation}
where the states $\left|k\right\rangle $ are the states $\left|0\right\rangle $,
$\left|\pm1\right\rangle $. The transfer matrix $F$ is obtained
as\begin{equation}
F=\left(\begin{array}{cccc}
1 & 0 & 0 & a\\
0 & -1 & 0 & 0\\
0 & 0 & -1 & 0\\
a & 0 & 0 & 1\end{array}\right)\label{21}\end{equation}
The eigenvalues are \begin{equation}
\varepsilon_{1}=1+a,\:\varepsilon_{2}=1-a,\:\varepsilon_{3}=-1,\:\varepsilon_{4}=-1\label{22}\end{equation}
The corresponding eigenvectors are\begin{equation}
\begin{array}{cc}
\left|e_{1}\right\rangle =\frac{1}{\sqrt{2}}\left(\begin{array}{c}
1\\
0\\
0\\
1\end{array}\right),\quad & \left|e_{2}\right\rangle =\frac{1}{\sqrt{2}}\left(\begin{array}{c}
-1\\
0\\
0\\
1\end{array}\right)\\
\left|e_{3}\right\rangle =\left(\begin{array}{c}
0\\
1\\
0\\
0\end{array}\right),\quad & \left|e_{4}\right\rangle =\left(\begin{array}{c}
0\\
0\\
1\\
0\end{array}\right)\end{array}\label{23}\end{equation}
From Eq. $(20)$,\begin{equation}
\rho(i)=\frac{\sum_{\alpha=1}^{4}\left\langle e_{\alpha}\right|F^{L-1}f\left|e_{\alpha}\right\rangle }{\sum_{\alpha=1}^{4}\left\langle e_{\alpha}\right|F^{L}\left|e_{\alpha}\right\rangle }\label{24}\end{equation}
The factor in the denominator takes care of the condition $Tr\,\rho=1$.
On taking the thermodynamic limit $L\rightarrow\infty$, we get \begin{equation}
\rho(i)=\varepsilon_{1}^{-1}\,\left\langle e_{1}\right|f\left|e_{1}\right\rangle \label{25}\end{equation}
In the $\left|0,\pm1\right\rangle $ basis, the reduced density matrix
becomes\begin{equation}
\rho(i)=\left(\begin{array}{ccc}
\frac{1}{1+a} & 0 & 0\\
0 & \frac{a}{2(1+a)} & 0\\
0 & 0 & \frac{a}{2(1+a)}\end{array}\right)\label{26}\end{equation}

The calculation of the two-site reduced density matrix $\rho(i,j)$
follows in the same manner. $\rho(i,j)$ is given by\begin{equation}
\rho(i,j)=Tr_{1,..L}^{i,j}\left|\psi_{0}\right\rangle \left\langle \psi_{0}\right|\label{27}\end{equation}
where the trace is taken over all the spins except the $i$-th and
$j$-th ones. \begin{equation}
\rho(i,j)=\frac{\sum_{\alpha=1}^{4}\left\langle e_{\alpha}\right|F^{i-1}f\, F^{j-i-1}f\, F^{L-j}\left|e_{\alpha}\right\rangle }{\sum_{\alpha=1}^{4}\left\langle e_{\alpha}\right|F^{L}\left|e_{\alpha}\right\rangle }\label{28}\end{equation}
In the thermodynamic limit $L\rightarrow\infty$, $\rho(i,j)$ reduces
to \begin{equation}
\rho(i,j)=\sum_{\alpha=1}^{4}\varepsilon_{\alpha}^{-2}\,\left(\frac{\varepsilon_{\alpha}}{\varepsilon_{1}}\right)^{n+1}\,\left\langle e_{1}\right|f\left|e_{\alpha}\right\rangle \left\langle e_{\alpha}\right|f\left|e_{1}\right\rangle \label{29}\end{equation}
where $n=|j-i|$.

The matrix $\rho(i,j)$ is a $9\times9$ matrix and defined in the
two-spin basis states $\left|lm\right\rangle $ with the ordering
\begin{equation}
\left|lm\right\rangle \equiv\left|11\right\rangle ,\left|10\right\rangle ,\left|01\right\rangle ,\left|1-1\right\rangle ,\left|-11\right\rangle ,\left|00\right\rangle ,\left|0-1\right\rangle ,\left|-10\right\rangle ,\left|-1-1\right\rangle \label{30}\end{equation}
The non-zero matrix elements, $b_{pq}$ $(p=1,...,9,\, q=1,...,9)$,
of $\rho(i,j)$ are : \begin{equation}
\begin{array}{c}
b_{11}=b_{99}=\frac{a^{2}}{4(1+a)^{2}}-\frac{a^{2}}{4(1-a^{2})}\left(\frac{1-a}{1+a}\right)^{n}\\
b_{22}=b_{33}=b_{77}=b_{88}=\frac{a}{2(1+a)^{2}}\\
b_{44}=b_{55}=\frac{a^{2}}{4(1+a)^{2}}+\frac{a^{2}}{4(1-a^{2})}\left(\frac{1-a}{1+a}\right)^{n}\\
b_{23}=b_{32}=b_{46}=b_{64}=b_{56}=b_{65}=b_{78}=b_{87}=\frac{a}{2(1+a)}\left(-\frac{1}{1+a}\right)^{n}\\
b_{66}=\frac{1}{(1+a)^{2}}\end{array}\label{31}\end{equation}
It is easy to check that $\rho(i,j)$ has a block-diagonal form.

\section*{III. ENTANGLEMENT MEASURES $\; S(i)$, $S(i,j)$, $G(2,n)$}

We now determine the entanglement content of the ground state $\left|\psi_{0}\right\rangle $
(Eq. $(9)$) of the Hamiltonian (Eq. $(5)$) using the entanglement
measures $S(i)$, $S(i,j)$, and $G(2,n)$. The calculations are carried
out for different values of the parameter $a$ in Eq. $(8)$. The
ultimate aim is to probe the special features, if any, of entanglement
in the vicinity of the QCP at $a=0$. From Eq. $(3)$ and $(26)$,
the one-site entanglement \begin{equation}
S(i)=\frac{1}{1+a}[(1+a)log_{2}\,(1+a)-a\, log_{2}\, a+a]\label{32}\end{equation}
Figure $1$ (top) shows the variation of $S(i)$ w.r.t. $a$. The
one-site entanglement has the maximum possible value $log_{2}\,3$.
This is attained at the AKLT point $a=2$. The VBS state is in this
case the exact ground state. In the VBS state, each spin-1 at a specific
lattice site can be considered as a symmetric combination of two spin-$\frac{1}{2}$'s
\cite{key-14}. In the VBS state, each spin-$\frac{1}{2}$ at a particular
lattice site forms a spin singlet with a spin-$\frac{1}{2}$ at a
neighbouring lattice site. $S(i)$ has the value zero at the QCP $a=0$.
Figure $1$ (bottom) shows the variation of $\frac{\partial S(i)}{\partial a}$
with the parameter $a$. The derivative diverges as the QCP is approached.
This is the expected behaviour at the QCP of a conventional QPT. In
the latter case, however, $S(i)$ has the maximum value at the QCP
\cite{key-3}.

From Eq. $(1)$ and $(31)$, the two-body entanglement $S(i,j)$ is
\begin{equation}
S(i,j)=-\sum_{i=1}^{9}\lambda_{i}\, log_{2}\lambda_{i}\label{33}\end{equation}
Where $\lambda_{i}$'s are the eigenvalues of the reduced density
matrix $\rho(i,j)$. These are given by\begin{equation}
\begin{array}{c}
\lambda_{1}=\lambda_{2}=\frac{a}{2(1+a)^{2}}-\frac{a}{2(1+a)}\left(-\frac{1}{1+a}\right)^{n}\\
\lambda_{3}=\lambda_{4}=\frac{a}{2(1+a)^{2}}+\frac{a}{2(1+a)}\left(-\frac{1}{1+a}\right)^{n}\\
\lambda_{5}=\frac{a^{2}}{4(1+a)^{2}}+\frac{a^{2}}{4(1-a^{2})}\left(\frac{1-a}{1+a}\right)^{n}\\
\lambda_{6}=\lambda_{7}=\frac{a^{2}}{4(1+a)^{2}}-\frac{a^{2}}{4(1-a^{2})}\left(\frac{1-a}{1+a}\right)^{n}\\
\lambda_{8}=\frac{1}{2}\left(\frac{a^{2}}{4(1+a)^{2}}+\frac{a^{2}}{4(1-a^{2})}\left(\frac{1-a}{1+a}\right)^{n}+\frac{1}{1+a^{2}}\right)-\frac{1}{2(1+a)}\\
\left(\frac{(a^{2}-4)^{2}}{16(1+a)^{2}}+2a^{2}\left(-\frac{1}{1+a}\right)^{2n}+\frac{a^{4}}{16(1-a)^{2}}\left(\frac{1-a}{1+a}\right)^{2n}+\frac{a^{2}(a^{2}-4)}{8(1-a^{2})}\left(\frac{1-a}{1+a}\right)^{2n}\right)^{\frac{1}{2}}\\
\lambda_{9}=\frac{1}{2}\left(\frac{a^{2}}{4(1+a)^{2}}+\frac{a^{2}}{4(1-a^{2})}\left(\frac{1-a}{1+a}\right)^{n}+\frac{1}{1+a^{2}}\right)+\frac{1}{2(1+a)}\\
\left(\frac{(a^{2}-4)^{2}}{16(1+a)^{2}}+2a^{2}\left(-\frac{1}{1+a}\right)^{2n}+\frac{a^{4}}{16(1-a)^{2}}\left(\frac{1-a}{1+a}\right)^{2n}+\frac{a^{2}(a^{2}-4)}{8(1-a^{2})}\left(\frac{1-a}{1+a}\right)^{2n}\right)^{\frac{1}{2}}\end{array}\label{34}\end{equation}
Knowing the reduced density matrix $\rho(i,j)$, the correlation functions
$\left\langle S_{i}^{\alpha}\, S_{j}^{\alpha}\right\rangle \;$ ($\alpha=x,y,z$)
can be calculated in the usual manner. One then recovers the expressions
in Eq. $(13)$ and $(14)$ ($r=n+1$, where $n=|j-i|$). Figure $2$
(top) shows the variation of $S(i,j)$ as a function of $a$ for $n=1000$.
Figure $2$ (bottom) shows the variation of the derivative $\frac{\partial S(i,j)}{\partial a}$
w.r.t. $a$ for the same value of $n$. The maximum of $S(i,j)$ is
at the AKLT point $a=2$ and has the value zero at $a=0$. For large
$n$, the derivative $\frac{\partial S(i,j)}{\partial a}$ diverges
near the QCP at $a=0$. The last fearure is characteristic of a conventional
QPT \cite{key-7}.

We next calculate the GGE $G(2,n)$ (Eq. $(4)$). This is easily done
as the matrix elements of the reduced density matrix (Eq. $(3)$)
are known. Figure $3$ (top) shows the variation of $G(2,n)$ versus
$a$ for $n=1000$. Figure $3$ (bottom) shows the plot of $\frac{\partial G(2,n)}{\partial a}$
against $a$. Again $G(2,n)$ has the maximum value at the AKLT point
and is zero at $a=0$. The derivative $\frac{\partial G(2,n)}{\partial a}$
does not diverge as $a\rightarrow0$ in contrast to the case of a
conventional QPT \cite{key-6}. The derivative, however, attains the
maximum value at the QCP $a=0$. Figure $4$ (top) shows the plots
of $S(i)$, $S(i,j)$, and $G(2,n)$ versus $a$ for $n=1000$. Figure
$4$ (bottom) shows the plots of the first derivatives of the same
quantities w.r.t. $a$ for $n=1000$. The plots are shown for comparing
the different entanglement measures.

We next determine the EL $\xi_{E}$ and its variation w.r.t. the parameter
$a$. We consider the entanglement measure $S(i,j)$ for this purpose.
Close to the QCP $a=0$ and in the limit of large $n$, one can write\begin{equation}
S(n=|j-i|)-S(\infty)\sim A_{e}\, e^{-\frac{n}{\xi_{E}}}\label{35}\end{equation}
The longitudinal and transverse correlation functions, $p_{n}^{z}=\left\langle S_{1}^{z}S_{n+1}^{z}\right\rangle $
and $p_{n}^{x}=\left\langle S_{1}^{x}S_{n+1}^{x}\right\rangle $ are
given by Eq. $(13)$ and $(14)$ with $r=n+1$. For $a<1$, $p_{n}^{z}$
decays faster than $p_{n}^{x}$ with $n$. The eigenvalues $\lambda_{i}$'s,
$i=1,...9$, can be expressed in terms of the correlation functions
$p_{n}^{z}$ and $p_{n}^{x}$. For large $n$, the contributions from
$p_{n}^{z}$ can be ignored. The eigenvalues now reduce to the expressions

\begin{equation}
\begin{array}{c}
\lambda_{1}=\lambda_{2}=\frac{a}{2(1+a)^{2}}-4\, p_{n}^{x}\\
\lambda_{3}=\lambda_{4}=\frac{a}{2(1+a)^{2}}+4\, p_{n}^{x}\\
\lambda_{5}=\frac{a^{2}}{4(1+a)^{2}}\\
\lambda_{6}=\lambda_{7}=\frac{a^{2}}{4(1+a)^{2}}\\
\lambda_{8}=\frac{1}{2}\left(\frac{a^{2}}{4(1+a)^{2}}+\frac{1}{1+a^{2}}\right)-\frac{1}{2}\left(\frac{(a^{2}-4)^{2}}{16(1+a)^{2}}+2\,\left(p_{n}^{x}\right)^{2}\right)^{\frac{1}{2}}\\
\lambda_{9}=\frac{1}{2}\left(\frac{a^{2}}{4(1+a)^{2}}+\frac{1}{1+a^{2}}\right)+\frac{1}{2}\left(\frac{(a^{2}-4)^{2}}{16(1+a)^{2}}+2\,\left(p_{n}^{x}\right)^{2}\right)^{\frac{1}{2}}\end{array}\label{36}\end{equation}
From Eq. $(1)$ and $(36)$, one ultimately arrives at the expressions\begin{equation}
S(n=|j-i|)-S(\infty)\sim A_{e}^{'}\,\left(p_{n}^{x}\right)^{2}\sim A_{e}\, e^{-\frac{n}{\xi_{E}}}\label{37}\end{equation}
The pre-factor $A_{e}=0$ at the QCP $\: a=0$ . The EL $\:\xi_{E}$
is given by \begin{equation}
\xi_{E}=\frac{\xi_{t}}{2}=\frac{1}{2\, ln(1+a)}\label{38}\end{equation}
where $\xi_{t}$ is the transverse correlation length (Eq. $(15)$).
In the case of the $S=\frac{1}{2}$ anisotropic XY model in a transverse
field, an expression similar to that in Eq. $(37)$ is obtained close
to the QCP in the limit of large $n$ \cite{key-7}. The pre-factor
in this case, however, does not vanish at the QCP but has a power-law
dependence on $n$. Figure $5$ shows the variation of $\xi_{E}$
w.r.t. $a$ based on the entanglement measure $S(i,j)$.

The total correlations, with both classical and quantum components,
between two sites $i$ and $j$ are quantified in terms of the quantum
mutual information \cite{key-21,key-22}\begin{equation}
I_{ij}=S(i)+S(j)-S(i,j)\label{39}\end{equation}
As explained in \cite{key-21}, a comparison of the singular behaviour
of $S(i)$ with that of $I_{ij}$ allows one to determine whether
two-point ($Q2$) or multipartite ($QS$) quantum correlations are
important in a QPT. Figure $6$ (top) shows a plot of $I_{ij}$ versus
$a$ for $n=1000$. Figure $6$ (bottom) shows the variation of $\frac{\partial I_{ij}}{\partial a}$
versus $a$ for $n=1000$. The derivative does not diverge as $a\rightarrow0$,
a behaviour distinct from that of $\frac{\partial S(i)}{\partial a}$
close to $a=0$. The difference in the singular behaviour of quantities
associated with $S(i)$ and $I_{ij}$ indicates that multipartite
quantum correlations are involved in the QPT. 

\begin{figure}
\begin{center}\includegraphics{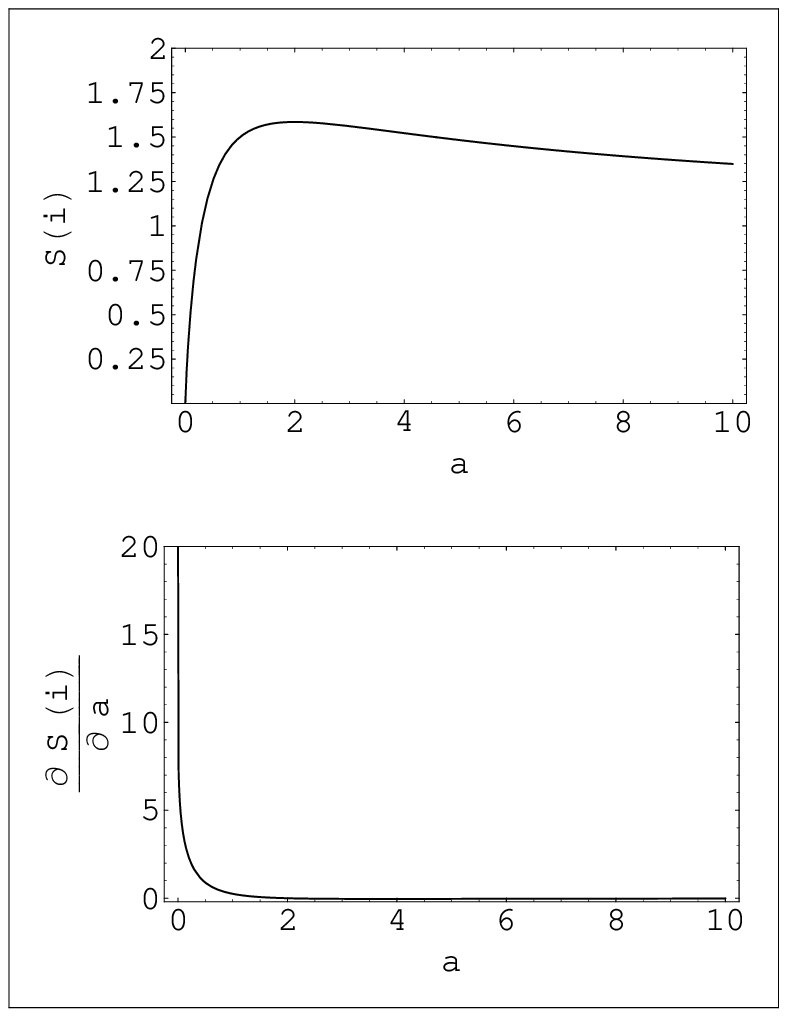}\end{center}

\textbf{FIG. 1:} Plot of $S(i)$ (top) and $\frac{\partial S(i)}{\partial a}$
(bottom) vs. $a$ .
\end{figure}

\begin{figure}
\begin{center}\includegraphics{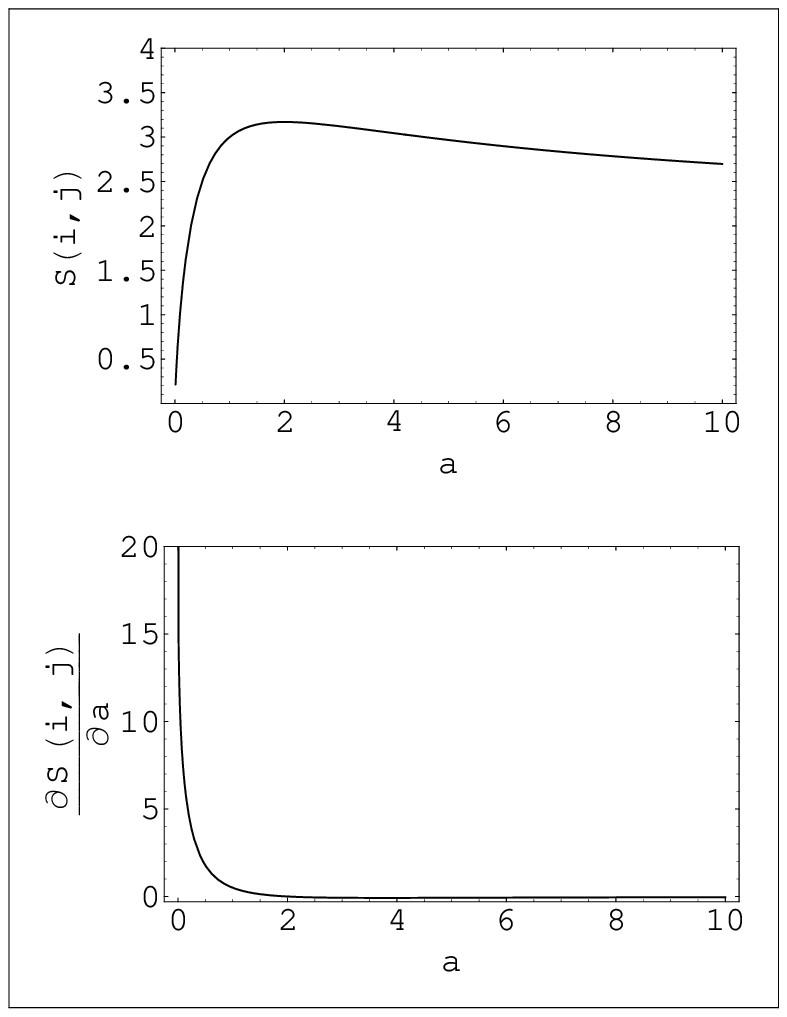}\end{center}

\textbf{FIG.} \textbf{2:} Plot of $S(i,j)$ (top) and $\frac{\partial S(i,j)}{\partial a}$
(bottom) as a function of $a$.
\end{figure}

\begin{figure}
\begin{center}\includegraphics{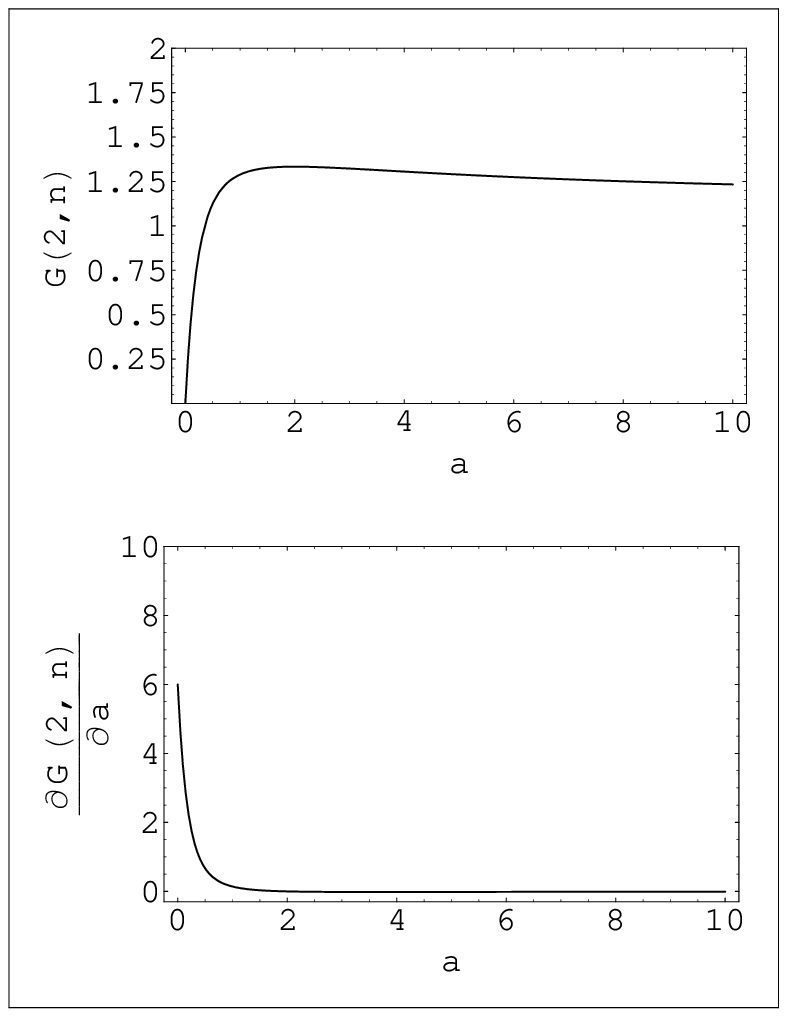}\end{center}

\textbf{FIG. 3:} Plot of $G(2,n)$ (top) and $\frac{\partial G(2,n)}{\partial a}$
(bottom) as a function of $a$.
\end{figure}

\begin{figure}
\begin{center}\includegraphics{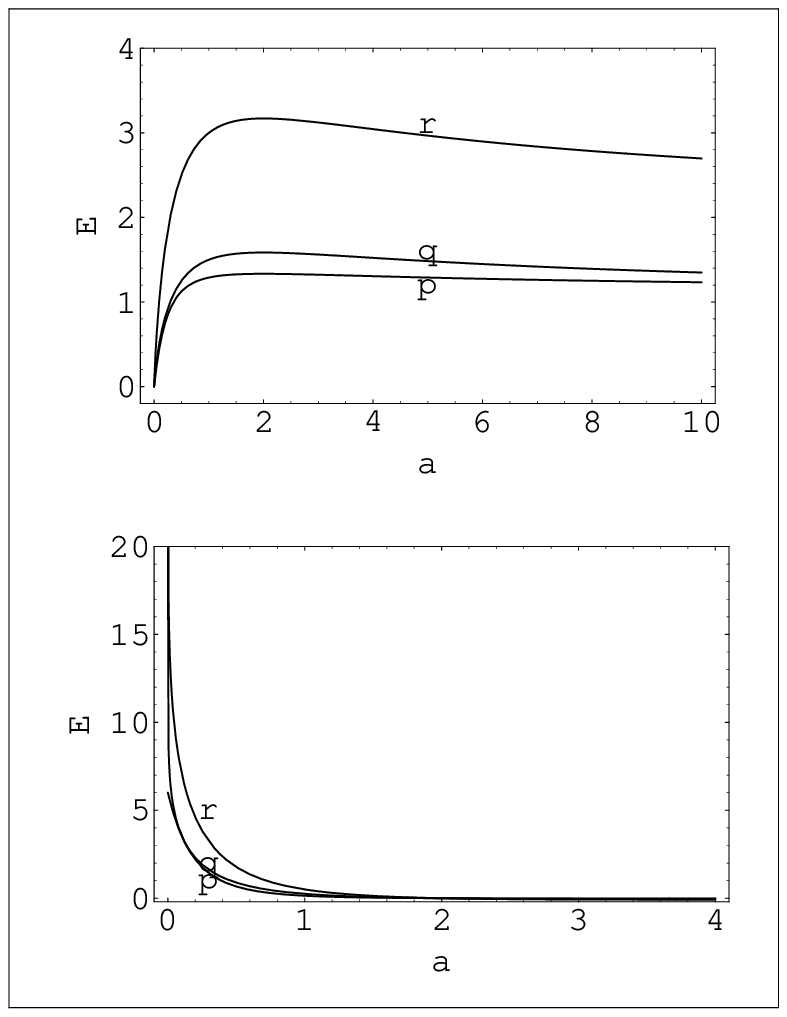}\end{center}

\textbf{FIG. 4:} Plots of $S(i)$ ($q$), $S(i,j)$ ($r$), and $G(2,n)$
($p$) (top) and the corresponding first derivatives w.r.t. $a$ (bottom)
as a function of $a$. $E$ represents the entanglement measure.
\end{figure}

\begin{figure}
\begin{center}\includegraphics{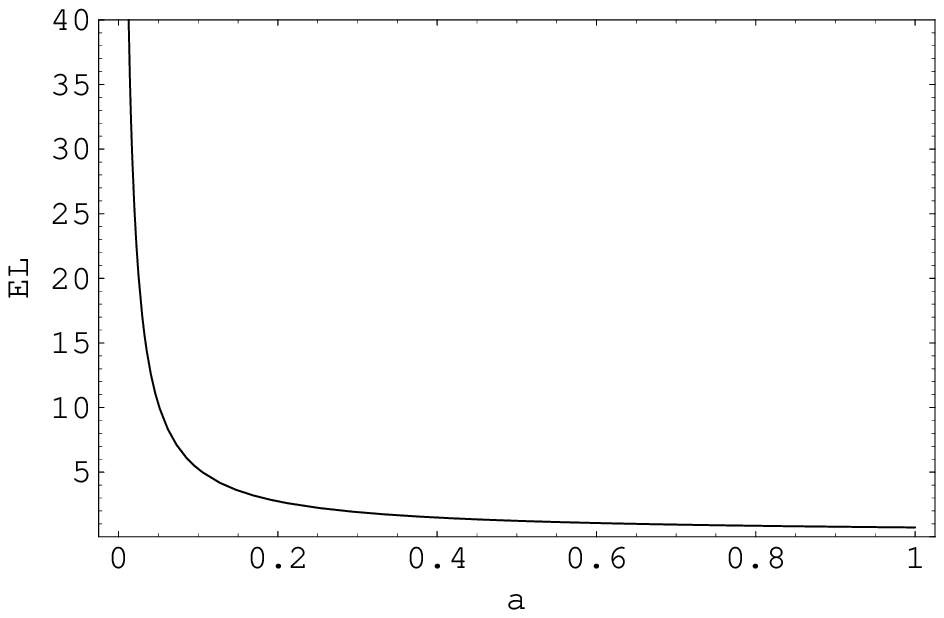}\end{center}

\textbf{FIG.} \textbf{5:} Plot of EL as a function of $a$.
\end{figure}

\begin{figure}
\begin{center}\includegraphics{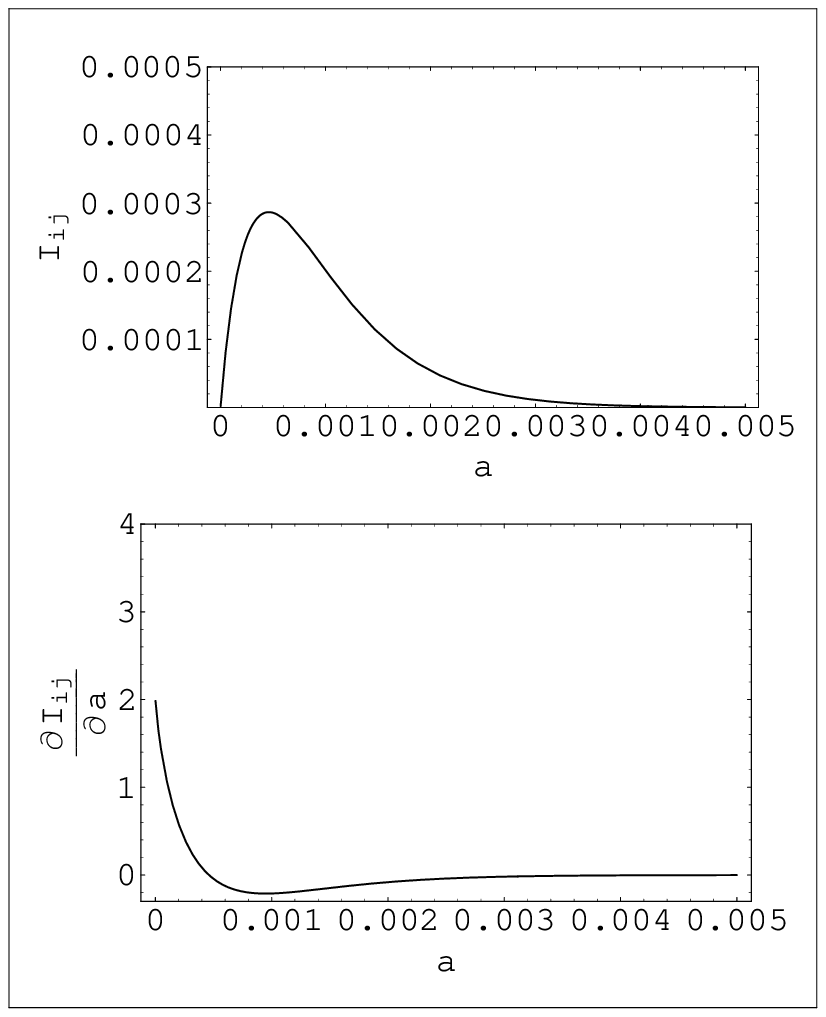}\end{center}

\textbf{FIG.} \textbf{6:} Plots of $I_{ij}$ (top) and $\frac{\partial I_{ij}}{\partial a}$
(bottom) as a function of $a$.
\end{figure}

\section*{IV. DISCUSSIONS}

The MP states provide exact representations of the ground states of
several spin models in low dimensions \cite{key-17}. The remarkable
features of such states arises from the fact that complicated many
body states have a simple factorized form. The simplicity in structure
makes the calculation of the ground state expectation values particularly
easy to perform. The spin-1 AKLT model is a well-known example of
a spin model in 1d the exact ground state of which (a VBS state) has
an MP representation. The AKLT model and the spin-1 Heisenberg AFM
chain belong to the same universality class \cite{key-23}. The insight
gained from the study of models in the MP formalism is expected to
be of relevance in understanding the properties of more physical systems.
The MP states also serve as good trial wave functions for standard
spin models. The MP representation lies at the heart of the powerful
density matrix renormalization group (DMRG) method and provides the
basis for several interesting developments in quantum information
\cite{key-24}.

Studies of the entanglement characteristics of the MP states have
begun only recently. The QPTs which occur in such states have characteristics
different from those of conventional QPTs. It is thus of considerable
interest to determine whether the entanglement content of MP states
develops special features close to a QCP. In this paper, we consider
a spin-$1$ model the exact ground state of which is of the MP form
over a wide range of parameter values. The model exhibits a novel
QPT in that the longitudinal and transverse correlation lengths diverge
as the QCP is approached but the correlation functions vanish at the
QCP. In a conventional QPT, the correlation functions have a power-law
decay at the QCP. In the spin-1 model, the divergence of correlation
lengths is accompanied by the excitation gap going to zero. The string
order parameter, which has a non-zero expectation value in the MP
state for $a>0$, vanishes at the QCP $a=0$. The distinctive signatures
indicate the appearance of a new phase. We study the entanglement
properties of the MP state for different values of the parameter $a$.
The measures used are $S(i)$ (one-site von Neumann entropy), $S(i,j)$
(two-body entanglement) and $G(2,n)$ (GGE). All the entanglement
measures have zero value at the QCP so that the ground state is disentangled
at that point. As seen from the different plots, figures $(1)$, $(2)$,
$(3)$, and $(4)$, the entanglement content, as measured by $E=$$S(i)$,
$S(i,j)$ and $G(2,n)$, has a slow variation w.r.t. $a$ for $a>2$.
At the AKLT point $a=2$, $E$ reaches its maximum value and as $a$
is reduced further, the magnitude of $E$ falls rapidly to approach
zero value at $a=0$. The study of conventional QPTs shows that $E$
is maximum at a QCP \cite{key-3,key-6,key-7}. Also, $\frac{\partial E}{\partial a}$
diverges as the QCP is approached. The EL, $\xi_{E}$, as calculated
from $S(i,j)$ and $G(2,n)$ for large $n$, also diverges with $\xi_{E}=\frac{\xi_{C}}{2}$
where $\xi_{C}$ is the usual correlation length. In the case of the
spin-1 model under consideration, $\frac{\partial E}{\partial a}$
diverges as $a\rightarrow0$ when $E=S(i)$ and $S(i,j)$. The EL,
$\xi_{E}$, as calculated from $S(i,j)$ in the large $n$ limit,
also diverges with $\xi_{E}=\frac{\xi_{C}}{2}$. One now has the interesting
situation that the entanglement content of the MP state decreases
as $a\rightarrow0$ but the entanglement is spread over larger distances.
The derivative $\frac{\partial G(2,n)}{\partial a}$, however, does
not diverge as $a\rightarrow0$ but attains a maximum value at the
QCP. The results can be understood by noting that in all the three
cases, $E=S(i),\: S(i,j)$ and $G(2,n)$, the reduced density matrices
$\rho(i)$ and $\rho(i,j)$ smoothly approach the forms associated
with pure states as the parameter $a$ tends to zero. The matrix elements
of the reduced density matrices do not develop non-analyticities in
the parameter region of interest. Thus, the energy density, calculated
from the reduced density matrix $\rho(i,j)$, does not develop a non-analyticity
at the QCP. The derivative $\frac{\partial G(2,n)}{\partial a}$ depends
upon the first derivatives of the matrix elements of $\rho(i,j)$.
Since the latter is analytic for $a\geq0$, $\frac{\partial G(2,n)}{\partial a}$
does not diverge in the whole parameter regime including the point
$a=0$. In the cases of the entanglement measures $E=S(i)$ and $S(i,j)$,
the derivative $\frac{\partial E}{\partial a}$ diverges as $a\rightarrow0$
due to the divergence of $log_{2}\, a$ in the same limit. The divergence
is thus due to the special form of the von Neumann entropy and occurs
for $n\geq1$. A recent work \cite{key-25} provides another example
of such a singularity. Though $\frac{\partial G(2,n)}{\partial a}$
does not diverge or become discontinuous at the QCP $a=0$, it attains
its maximum value at the point. The first derivative of the string
order parameter w.r.t. $a$ also attains its maximum value at $a=0$
though the order parameter itself vanishes at the point. Figure $(4)$
shows that amongst the three entanglement measures $E=$$S(i)$, $S(i,j)$
and $G(2,n)$, used in this study to obtain a quantitative estimate
of the entanglement content of the MP ground state, the measure $S(i,j)$
yields the largest value of the entanglement at different values of
$a$. The difference in the singular behaviour of the measures $S(i)$
and the mutual information entropy $I_{ij}$ as $a\rightarrow0$ indicates
that multipartite quantum correlations are involved in the QPT. In
summary, the present study identifies the entanglement characteristics
of the MP ground state of a spin-1 model close to the critical point
$a=0$. The features are distinct from those associated with conventional
QPTs. Several spin models are known for which the MP states are the
exact ground states \cite{key-17}. Some of these models have interesting
phase diagrams exhibiting both first and second order phase transitions.
It will be of interest to extend the present study to other spin models
(both $S=\frac{1}{2}$ and $1$) in order to identify the universal
characteristics of QPTs in MP states.

\section*{ACKNOWLEDGMENT}

A. T. is supported by the Council of Scientific and Industrial Research,
India, under Grant No. 9/15 (306)/ 2004-EMR-I.


\begin{thebibliography}{10}
\bibitem{key-1}S. Sachdev, Quantum Phase Transitions (Cambridge University Press,
Cambridge, 2000); S. L. Sondhi et al., Rev. Mod. Phys. 69, 315 (1997).
\bibitem{key-2}A. Osterloh, L. Amico, G. Falci and R. Fazio, Nature 416, 608 (2002).
\bibitem{key-3}T. J. Osborne and M. A. Nielsen, Phys. Rev. A 66, 032110 (2002).
\bibitem{key-4}G. Vidal, J. I. Latorre, E. Rico and A. Kitaev, Phys. Rev. Lett. 90,
0279011 (2004).
\bibitem{key-5}L. -A. Wu, M. S. Sarandy and D. A. Lidar, Phys. Rev. Lett 93, 250404
(2004).
\bibitem{key-6}T. R. de Oliveira, G. Rigolin, M. C. de Oliveira and E. Miranda, Phys.
Rev. Lett. 97, 170401 (2006).
\bibitem{key-7}H.- D. Chen, cond-mat/0606126.
\bibitem{key-8}I. Bose and E. Chattopadhyay, Phys. Rev. A 66, 062320 (2002).
\bibitem{key-9}F. C. Alcaraz, A. Saguia and M. S. Sarandy, Phys. Rev. A 70, 032333
(2004).
\bibitem{key-10}J. Vidal, R. Mosseri and J. Dukelsky, Phys. Rev. A 69, 054101 (2004).
\bibitem{key-11}F. Verstraete, M. Popp and J. I. Cirac, Phys. Rev. Lett. 92, 027901
(2004).
\bibitem{key-12}H. Fan, V. Korepin and V. Roychoudhury, Phys. Rev. Lett. 93, 227203
(2004).
\bibitem{key-13}F. Verstraete, M. A. Martin-Delgado and J. I. Cirac, Phys. Rev. Lett.
92, 087201 (2004).
\bibitem{key-14}I. Affleck, T. Kennedy, E. H. Lieb and H. Tasaki, Phys. Rev. Lett.
59, 779 (1987); Commun. Math. Phys. 115, 447 (1988). 
\bibitem{key-15}L. Campos Venuti and M. Roncaglia, Phys. Rev. Lett. 94, 207207 (2005).
\bibitem{key-16}A. Kl\"{u}mper, A. Schadschneider and J. Zittartz, Z. Phys. B 87,
281-287 (1992).
\bibitem{key-17}A. K. Kolezhuk and H. J. Mikeska, Int. J. Mod. Phys. B 12, 2325 (1998)
and references therein.
\bibitem{key-18}M. Fannes, B. Nachtergaele and R. F. Werner, Europhys. Lett. 10, 633
(1989); Commun. Math. Phys. 144, 443 (1992).
\bibitem{key-19}M. M. Wolf, G. Ortiz, F. Verstraete and J. I. Cirac, Phys. Rev. Lett.
97, 110403 (2006).
\bibitem{key-20}A. Kl\"{u}mper, A. Schadschneider and J. Zittartz, Europhys. Lett.
24, 293 (1993).
\bibitem{key-21}A. Anfossi, P. Giorda, A. Montorsi and F. Traversa, Phys. Rev. Lett.
95, 056402 (2005).
\bibitem{key-22}A. Anfossi, P. Giorda, A. Montorsi and F. Traversa, cond-mat/0611091.
\bibitem{key-23}S. Brehmer, H. -J. Mikeska and U. Neugebauer, J. Phys.: Condens. Matter
8, 7161 (1996).
\bibitem{key-24}D. P\'{e}rez-Garc\'{i}a, F. Verstraete, M. M. Wolf and J. I. Cirac,
quant-ph/0608197.
\bibitem{key-25}M. Cozzini, R. Ionicioiu and P. Zanardi, Cond-mat/0611727.\end{thebibliography}
\end{document}